# Demystifying Quantum Blockchain for Healthcare

Keshav Kaushik[a,*], Adarsh Kumar[b]

[a,b]*School of Computer Science, University of Petroleum and Energy Studies, Dehradun, Uttarakhand, India*
*officialkeshavkaushik@gmail.com, adarsh.kumar@ddn.upes.ac.in*

**Abstract**

The application of blockchain technology can be beneficial in the field of healthcare as well as in the fight against the COVID-19 epidemic. In this work, the importance of blockchain is analyzed and it is observed that blockchain technology and the processes associated with it will be utilised in the healthcare systems of the future for data acquisition from sensors, automatic patient monitoring, and secure data storage. This technology substantially simplifies the process of carrying out operations because it can store a substantial quantity of data in a dispersed and secure manner, as well as enable access whenever and wherever it is required to do so. With the assistance of quantum blockchain, the benefits of quantum computing, such as the capability to acquire thermal imaging based on quantum computing and the speed with which patients may be located and monitored, can all be exploited to their full potential. Quantum blockchain is another tool that can be utilised to maintain the confidentiality, authenticity, and accessibility of data records. The processing of medical records could potentially benefit from greater speed and privacy if it combines quantum computing and blockchain technology. The authors of this paper investigate the possible benefits and applications of blockchain and quantum technologies in the field of medicine, pharmacy and healthcare systems. In this context, this work explored and compared quantum technologies and blockchain-based technologies in conjunction with other cutting-edge information and communications technologies such as ratification intelligence, machine learning, drones, and so on.

## 1. Introduction to Quantum Computing

Quantum computing (QC) is a promising computing approach based on quantum physics and its extraordinary events. It's a beautiful synthesis of mathematics, physics, computer science, and computational modelling. It achieves tremendous processing capacity, low energy consumption, and exponential speed above traditional computers by regulating the behaviour of tiny physical things such as atoms, electrons, photons, and other minuscule particles. Considering quantum theory is a broader paradigm of science than classical physics, it contributes to a more general framework of computing, quantum computing, that can address problems that classical computing can't. Unlike regular computers, which use binary bits 0 and 1 to store and process data individually, QC uses their quantum bits, often known as 'Qubits.' 'Quantum Computers' are computers that use quantum computing. QC can quickly penetrate today's encryption techniques, but the most incredible supercomputer presently available takes thousands of years. While QC will be competent in deciphering several of today's encryption techniques, it is believed that they will build hack-proof replacements. Transistors, logic gates, and Integrated Circuits cannot be used in such small computers. Therefore, atoms, protons, electrons, and ions are used as bits and their rotation and state metadata. They may be layered to make new combinations. Consequently, they may run parallel and efficiently employ memory, increasing their power. QC is the only computing paradigm that defies the Church-Turing thesis, permitting QC to take advantage of the available systems several times greater.

The quantum bit or qubit's central element of quantum theory depicts elementary particles such as atoms, electrons, and other subatomic particles as computer memory while their regulatory systems act as computer processors. It can have a value of 0, 1, or both simultaneously. It has a million times the processing power of today's most advanced and powerful. In engineering, producing and managing qubits is a considerable task. The quantum computer's computing strength comes from its digital and analog nature. Quantum gates have no distortion limit due to their analog nature, yet their digital nature gives a standard for recovering from this significant flaw. As a result, the logic gates and representations used in classical computing are useless in quantum computing. Purely classical computing principles can be used in quantum computing. However, this computation requires a unique way to avoid processing variances and any form of noise. It also requires its technique for debugging issues and dealing with design flaws.

There are three essential properties [1] of QC – superposition, interference, and entanglement. In quantum computing, superposition denotes a quantum system's capacity to exist simultaneously in two distinct places or configurations. It allows incredible parallel processing with high speed and is quite different from its classical counterparts, with binary restrictions. The QC system stores information in two states at the same time. In QC, interference is comparable to wave interference in traditional physics. Two waves strike in a single environment, leading to wave interference. However, suppose the waves are aligned in the same direction. In that case, it generates standing waves with respective amplitudes added collectively, referred to as constructive interference, or a consequent wave with their amplitudes wiped out, known as destructive interference. Depending on what type of interference, the net wave might be larger or less than the original wave. One of the essential aspects of quantum computing is entanglement. It refers to the close relationship between two quantum particles or qubits. Regardless if they are separated by huge distances, like at opposing ends of the Universe, qubits are linked in a flawless immediate relationship. They are intertwined or characterized by one another.

There are numerous quantum computing applications, but some prominent ones [2] are highlighted in figure 1. The main applications of quantum computing involve – Cybersecurity [3] [4], healthcare [5], artificial intelligence [6], financial modelling, logistics optimization, and weather forecasting. Because of the growing number of cyber-attacks that arise along the way all over the world, the internet security [7] [8] environment has become highly susceptible. Even though businesses are implementing the appropriate security frameworks, the procedure for traditional digital computers has become intimidating and unworkable. Figure 1 explains about the multiple applications of quantum computing, some of them are as follows:

- Cybersecurity - Quantum computers on a large scale will considerably increase computational capability, opening up new possibilities for enhancing cybersecurity. Quantum-period cybersecurity will be able to identify and block cyberattacks from that era before they do damage. But it could end up being a double-edged sword since quantum computing might also open up new vulnerabilities, such the capacity to swiftly solve the challenging mathematical puzzles that form the basis of some types of encryption. Businesses and other groups may start getting ready now even if post-quantum cryptography guidelines are still being developed.
- Healthcare - Combining quantum and classical computing in the healthcare industry is anticipated to offer significant benefits that classical computing alone cannot provide. A new style of understanding, a highly sought-after set of talents, unique IT architectures, and innovative business strategies are all required for quantum computing. Additionally, the technology directly affects security. Considering the sector's obligations and difficulties with regard to data privacy, security is a topic of special concern for the healthcare industry.
- Artificial Intelligence - AI and quantum computing are both game-changing technologies, and for artificial intelligence to make substantial strides [9], quantum computing is a must. Artificial intelligence is constrained by the computing power of conventional computers, while producing useful applications on them. Artificial intelligence may benefit from a compute boost from quantum computing, allowing it to handle more challenging issues in a variety of commercial and scientific domains.
- Financial Modelling - Financial firms that can use quantum computing will probably gain a lot from it. They will be better equipped to assess big or unstructured data collections, in particular. By making offers that are more timely or relevant, for instance, banks might make better judgments and provide better customer service. Where algorithms are driven by real-time data streams, such as real-time share prices, which contain significant random noise, quantum computers are showing promise.
- Logistics Optimization - Numerous benefits from quantum computing may be realised in the logistics industry. Current CPUs would be complemented by quantum computers, speeding up gadgets using machine learning and AI. Quantum supercomputers would have a significant impact on route planning in logistics. Utilizing quantum computing would improve the utilisation of warehouse modeling by examining all feasible routing choices and selecting the most effective one while accounting for all factors.
- Weather Forecasting - On a local and a larger scale, quantum computing can help weather forecasting for more sophisticated and precise warning of catastrophic weather occurrences, possibly saving lives and lowering yearly damage to property. Beyond weather forecasting, keep up with the 1QBit blog and follow us on social media to learn more about the status of quantum computing and its growing effect on a range of sectors. By handling enormous amounts of data with numerous variables efficiently and quickly utilising the computing power of qubits, and applying quantum-inspired optimization algorithms, quantum computing has the potential to advance conventional numerical methods to enhance tracking and forecasts of weather conditions.

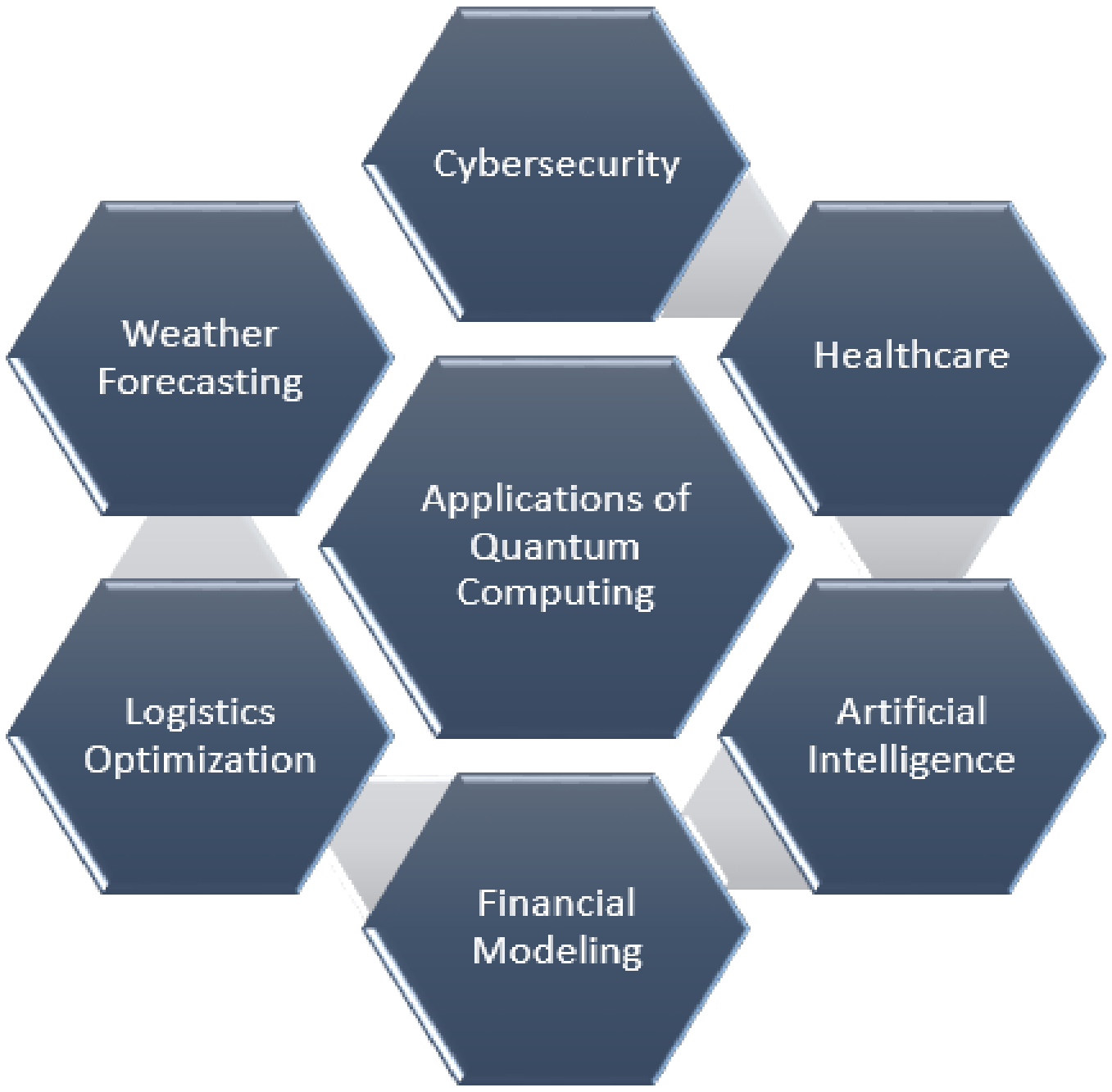

Fig. 1. Applications of Quantum Computing

Pharmaceutical formulation and construction are the most demanding tasks in QC. Typically, drugs are created through trial and error, which is pricey, dangerous, and time-consuming. QC, according to studies, might be a beneficial tool for investigating drugs and their impacts on people, potentially saving drug companies a lot of time and resources. Machine learning and artificial intelligence are two of the most critical issues today since new technologies have infiltrated nearly every aspect of human existence. QC can reduce its time to solve complex problems that might otherwise take years on conventional devices. Since accounting professionals manage vast sums, even a minor change in the expected return can significantly impact. Another possible use is algorithmic investing, wherein a computer performs complex procedures to automatically initiate share trades depending on market circumstances, which is beneficial, especially in high-volume transactions. Quantum annealing is a cutting-edge, efficient algorithm that might surpass traditional computers. On the other hand, Ubiquitous QC is prepared to address any computing issue but is not yet commercially accessible.

The paper is arranged into nine major sections, in the first section, the quantum computing is introduced followed by quantum blockchain in section two. Introduction of Quantum Blockchain in healthcare is discussed in section three, whereas advanced Quantum-integrated technologies like Quantum Drones, Quantum Satellites, Quantum AI, Quantum ML is highlighted in section four. Furthermore, Photonic Quantum Computing for Healthcare is added in section five. Quantum Gates and Circuits for Healthcare are highlighted in section six, and Quantum Algorithms for Healthcare is discussed in section seven. Quantum Simulation and Healthcare, comparative analysis of quantum computing-based approaches for healthcare are discussed in section eight and nine respectively. Finally, conclusion and future directions are discussed in section ten.

## 2. Introduction to Quantum Blockchain

The blockchain is a decentralized distributed hyperledger that stores and shares copies of all activities or digital events. The plurality of involved parties confirms each transaction. It holds every single transaction record. Blockchain- The groundbreaking technology transforming many sectors was mysteriously launched into the market with its first advanced form, Bitcoin. Bitcoin is a type of cryptocurrency that may be used to replace traditional money for transacting. Blockchain is the advanced technology that has led to the rise of cryptocurrencies. Quantum blockchain (QB) is encrypted, distributed, and decentralized based on quantum information theory and QC. The data will not be altered after being recorded on the QB. As QC and quantum communication theory has advanced in recent years, more academics have turned their attention to QB exploration. QB Technologies has launched an ambitious research, development, and investment programme in the fast-paced Blockchain Technology field, including cryptocurrency mining and other sophisticated blockchain applications. Quantum's research and development will be focused on cryptography, combining the most sophisticated implementation techniques and functions with quantum computing technologies and AI deep learning to provide a new and innovative perspective on blockchain technology.

The cryptocurrency business began to increase, with a market capitalization of $90 billion last month. Whatever happens next, cryptocurrencies will play an increasingly significant role in the world financial sector. The most challenging aspect of digital cash is ensuring that everybody uses it legitimately. Moreover, it appears that blockchain technology offers a promising alternative. This ensures honesty by employing cryptographic algorithms commonly considered impenetrable, barring brute force assaults. In this paper [10], the authors examined recent breakthroughs in QB and briefly discussed its benefits over the traditional blockchain. The QB's architecture and structure are discussed. Integrating quantum technology into a specific area of the broader blockchain is then shown.

Furthermore, the benefits of QB over the traditional blockchain are discussed, as well as its future development potential. This paper [10] demonstrates that time entanglement, rather than spatial entanglement, offers a critical quantum benefit. All of the system's constituent parts have been experimentally realized. Additionally, our encoding approach can potentially influence the past in a non-classical way. The paper [11] explores the fundamental problems, hazards, and benefits of putting these technologies into practice from a future perspective. We wrap up our study with an overview of the field's significant gaps, methodological issues, and suggestions for future research.

A blockchain is a distributed hyperledger that is cryptographically safeguarded from harmful changes. While blockchain systems promise a broad array of applications, they rely on digital certificates, which are susceptible to quantum computer assaults. Although to a lesser degree, the same holds for cryptographic algorithms used to prepare new blocks, implying that parties having access to quantum processing may have a square edge in obtaining mining rewards. Researchers [12] present an exploratory implementation of a quantum-safe blockchain platform for cognitive person identification that leverages quantum key distribution via an urban fibre network as a viable answer to the quantum era blockchain dilemma. The existing research developments on quantum computing, quantum-safe computing, or post-quantum cryptography, which are essential to quantum connections, are discussed in this paper [13], as well as the benefits involved, constraints, future breakthroughs, and research issues related to quantum technologies, as drones, and their systems. This research also includes constructing a categorization system for quantum-related disciplines depending on the rationale of their comprehension and study of each of these disciplines.

This chapter [14] surveys the topic of blockchain systems. The existing systems' security is predicated on computational outcome expectancy, and many mainstream encryption methods are susceptible to the arrival of full-fledged quantum systems. The authors [15] of this study proposed a unique negotiating process to set the authenticity of a block and allocate a new block in the blockchain infrastructure. Mediation processes based on an expanded probability environment are used to achieve block verification and allocation. A unique QB system is suggested in this study [16] to improve blockchain safety. First, the authors offered a description of QB and discussed its development. In particular, the benefits are outlined in this paper. Second, we created a new cryptocurrency called quantum coin based on the quantum no-cloning theorem. Furthermore, we build a unique QB method using quantum entanglement and DPoS.

A brief overview of quantum cryptography and the blockchain system is provided in the paper [17]. The foundation and inspiration for fast and secure online communications network-based methods are discussed. The concept is briefly described to develop a network-based data exchange application using QB technology. QB technology's primary purpose in transmitting data is to create long-term fast, stable network connectivity.

## 3. Importance of Quantum Blockchain for Healthcare

Insurance companies and digitally recorded healthcare datasets can assist society in reducing the healthcare ecosystem's high degree of complexity and cost. The general data protection legislation gives data owners the right to know how their data is maintained and used. Nevertheless, healthcare data is transmitted over an open channel, namely the Internet, allowing hackers to carry out nefarious actions such as breaching sensitive data, altering stored data, etc. As a result, maintaining the security and anonymity of participants is a problematic issue for traditional medical systems [18]. Blockchain has evolved as a platform that enhances today's healthcare system's efficiency while ensuring all parties' privacy and security. The authors [19] examine several security architectures used to safeguard health records and implement QC to the typical encryption system in this article, which is inspired by these findings. Then, for healthcare, we offer a blockchain-based architecture that allows individuals to access information from the database depending on their stated roles.

Electronic Medical Records (EMRs) to improve the performance and dependability of health care have become a common occurrence as medical knowledge advances. However, EMRs are housed separately in medicine and healthcare institutes [20], posing sharing issues. In addition, extremely sensitive EMRs are vulnerable to tampering and abuse, creating privacy and security risks. From a philosophical city standpoint, the latest advancements of Healthcare 4.0, which uses IoT elements [21] and cloud services to access medical procedures remotely, have piqued the researcher's interest. Frequent medical data monitoring, consolidation, data transfer, information sharing, and data management were the core components of Healthcare 4.0. Protecting patients' sensitive and confidential data from hackers presents various obstacles. As a result, storing, retrieving, and transmitting patient medical data on the cloud necessitates extra security precautions to ensure that authorized user elements of E-healthcare systems do not compromise it. Several cryptographic techniques have been created to provide secure health information storage, transmission, and retrieval in cloud service providers. Table 1 provides the comparative analysis of related work in the domain of Quantum Blockchain.

Table 1. Comparative Analysis of related work in the area of Quantum Blockchain

| Author | Year | A | B | C | D | E | Major Findings |
|---|---|---|---|---|---|---|---|
| Chen et al. [22] | 2022 | ✓ | ✓ | ✓ | ✓ | × | The authors developed an Anti-Quantum Attribute-based Authentication for Protected EMRs Sharing with Blockchain to address the issues. |
| Mirtskhulava et al. [23] | 2021 | ✓ | ✓ | ✓ | ✓ | × | Blockchains are seen as a critical answer for several 5G challenges, including mobile IoT security and the exchange of EHR. The researchers looked at Hashing-based Post-Quantum Authentication as a way to improve blockchain safety. |
| Mahajan et al. [24] | 2022 | × | ✓ | × | ✓ | × | This article thoroughly examines contemporary blockchain-based medical data security solutions, both with and without cloud computing. In this work, we use blockchain to build and analyze several strategies. The study gaps, problems, and future roadmap are the findings of this article, as per the research investigations, which promote rising Healthcare 4.0 innovation. |
| Azzaoui et al. [25] | 2022 | ✓ | ✓ | ✓ | ✓ | × | This study provides a Quantum Cloud-as-a-Service for sophisticated Smart Healthcare calculations that are efficient, scalable, and secure. What sets us distinct is our usage of Quantum Terminal Machines and Blockchain technology to increase the feasibility and anonymity of the proposed model. |
| Fernández-Caramès et al. [26] | 2020 | ✓ | ✓ | ✓ | × | × | This paper first examines the present state of post-quantum cryptographic protocols and how they may be used for blockchains and distributed ledger technologies. Prospective blockchain researchers and engineers will benefit from this paper's comprehensive perspective and practical guidance on post-QB security. |
| Gupta et al. [27] | 2022 | ✓ | ✓ | ✓ | × | ✓ | The authors presented a strategy to withstand a quantum assault by employing lattice cryptography. Furthermore, a trustworthy blockchain system is demonstrated to ensure the trustworthiness of automobiles in batch data verification. |
| Cai et al. [28] | 2019 | ✓ | ✓ | ✓ | × | × | The suggested quantum trademark strategies rely on quantum entanglement characteristics that a solitary signer or multiple signers may use and will enhance the stability of blockchain-based smart contracts against quantum attacks while maintaining a lightweight structure and eliminating the need for a reliable party or arbitrary organization. |
| Iovane [29] | 2021 | ✓ | ✓ | ✓ | × | × | In this research, we present a unique negotiation technique for fixing the validity of a transaction and allocating a new block in blockchain architecture. Negotiation processes based on an expanded likelihood framework are used to achieve block verification and allocation. |
| Li et al. [30] | 2018 | ✓ | ✓ | ✓ | × | × | The authors of this study provide an analysis of the contemporary blockchain networks' weaknesses to a quantum attacker and some prospective post-quantum mitigating strategies. |
| Gao et al. [31] | 2018 | ✓ | ✓ | ✓ | × | × | The authors presented a lattice problem-based signature system. The lattice-based delegating method establishes secret keys, while the preimage survey technique authenticates communications. |

A: Quantum Computing B: Blockchain C: Quantum Blockchain D: Healthcare E: Quantum Drones

## 4. Advanced Quantum-integrated technologies like Quantum Drones, Quantum Satellites, Quantum AI, Quantum ML

Exploring ways to employ quantum computing to speed up the implementation of conventional machine learning algorithms is a crucial objective of QML research. QML is also known as quantum-assisted machine learning or quantum-enhanced machine learning. As quantum computing progresses beyond being a developing and primarily theoretical technology, data scientists and others are becoming increasingly familiar with the application of quantum computing for machine learning and other operations. Nowadays, quantum computing is being tested in real-world settings through a few pilot programmes at major IT firms. The outcome could lead to more real-world and less theoretical quantum computing systems.

For the benefit of those who aren't aware, a "drone" is a kind of aircraft that may be controlled by someone other than the pilot but which flies in a manner analogous to that of an aeroplane or pilot. It is common to practice using this term while discussing automobiles. Unmanned aerial vehicles/drones and piloted aircraft may be differentiated from one another by the presence of a human pilot. It is physically impossible for an aeroplane to take off without a person at the controls (the autopilot mode is not distinguishable). Quantum AI and drones can help collect imaging in thermal space, keep an eye on things, and conduct surveillance are some of the uses of this technology. Other possible services include making adjustments to systems and distributing medicine or meals, sanitization and disaster relief force. Quantum drone-based systems with quantum satellites might be significantly improved by combining quantum-based data transfer, social distance measuring, and data collection using AI-based recognition. Quantum drones are useful in environment cleaning [8] and many more applications when enabled with quantum satellites, quantum AI and quantum ML technologies.

A theoretical area called quantum machine learning (QML) is only beginning to take shape. It sits where quantum computing and machine learning converge. Quantum Machine Learning's primary objective is to accelerate processes by integrating machine learning with what we've learned about quantum computing. QML theory adopts concepts from conventional machine learning theory and approaches quantum computing through those frameworks. For simulating quantum machine learning, Xanadu offers PennyLane as open-source software. It blends traditional machine learning software with hardware and quantum simulators. PennyLane supports a broad range of machine learning frameworks and a developing ecosystem of quantum hardware. Despite the community's ongoing efforts to develop fault-tolerant quantum theory, PennyLane and hardware options enable corporate clients to begin utilizing quantum computing right away.

Deep learning and quantum computing can be used together to speed up neural network training. Using this technique, we may accomplish fundamental optimization and create a new paradigm for deep learning. We can reproduce classical deep learning methods on a real, physical quantum computer. As more neurons are added, the computational complexity rises when multi-layer perceptron topologies are used. Performance may be enhanced by using specialized GPU clusters, which also considerably cuts down on training time. Indeed it, though, will rise in comparison to quantum computers.

The hardware of quantum computers is intended to imitate brain networks rather than the software found in traditional computers. Here, a qubit takes on the role of a neuron, the fundamental building block of a neural network. As a result, a quantum system with qubits may perform the function of a neural network and be utilized for deep learning applications at a rate that is faster than any traditional machine learning technique. In other terms, quantum machines have the potential to improve our quality of life and, when properly used, can remove many obstacles to the way we can improve machine learning algorithms.

## 5. Photonic Quantum Computing for Healthcare

The next major transformation in the medical industry will be led by quantum computing [32]. This technology has several benefits across various industries, particularly in those that impact the health sector. The fusion of quantum theory and quantum technology allows for the processing of massive amounts of data, the creation of simulations, the development of specialized medications, and the molecular manipulation of organs using nanotechnology, artificial intelligence, and particularly quantum computing. A new scientific subject called photonic quantum information has emerged due to recent technological advancements in synthesizing, managing, and sensing individual single photons. This development involves the creation of single photon switching, functional photonic quantum circuits, and innovative optical metrology that goes beyond the capabilities of conventional optics. Although photonics [33] presents unique benefits as a platform for quantum information processing, it faces significant scaling difficulties.

While deterministic methods require an unreasonably high number of identical quantum emitters to construct large quantum circuits, nondeterministic techniques incur enormous resource overheads.

We need to make and control a lot of qubits in order to build a working quantum computer. Although significant advancements have been achieved utilizing trapped ions, superconducting circuitry, and several other technologies, this accomplishment has proven to be challenging. Scalability, or the capacity to combine numerous qubits, is constrained since each qubit in a multi-qubit system gradually loses its quantum features. Decoherence is a phenomenon that arises from interactions between the qubits and their environment. Utilizing photons is one technique to generate scalable structures. We might be onto something here since photons' quantum states are more resistant to decoherence.

If the Wigner function of a photonic system is a two-dimensional Gaussian function, the system is indeed in a Gaussian state. An operation that would take an ordinary supercomputer more than 9,000 years to complete may be completed in only 36 microseconds by a new photonic quantum computer. The brand-new system, called Borealis, is the first quantum computer developed by a startup to exhibit such a "quantum edge" over conventional computers. Borealis is the first computer with a quantum advantage made publicly accessible through the cloud. Theoretically, quantum computers can acquire a quantum edge that permits them to discover solutions to issues that conventional computers have never been able to.

Compared to previous approaches, photonic devices take a very different technique to quantum computing. Even though the creation of qubits still poses a problem, recent theoretical and practical advancements have enhanced its reputation as a scalable solution. We can integrate qubits into photonic states in a wide range of ways, which leaves a lot of opportunity for inventiveness and pave the way for future research and technological advances.

## 6. Quantum Gates and Circuits for Healthcare

The state vectors are used by quantum gates, which are analogous to logic gates. The functions of quantum gates are described by Alice and Bob, who also show how they affect the states that describe a single qubit. The NOT gate, the Pauli gates, and the Hadamard gate are some examples of these gates. Even abstract characters and matrices that aggregate state vectors are used to express the gate functions. Additionally, Alice and Bob provide a brief introduction to a few of the mathematical properties of quantum gates and what occurs when many gates work sequentially on the state, denoting a qubit, a property that frequently recurs in quantum computing. Cardy is interested in how quantum measurements and quantum gates are connected. Alice and Bob make the critical distinctions between measurement tools and quantum gates. These distinctions hint at the substantial conceptual gap between quantum and conventional physics.

Doctors will be able to include a great number of cross-functional data sets into their physician's risk factor thanks to quantum's capacity to compute at scale. Thanks to quantum computing, a better match between the procedure and the patient will be ensured by employing additional points of reference when choosing clinical trial participants. Reliability and promptness in diagnosis and treatment are now essential for providing high-quality care. The use of quantum computing might lead to unparalleled processing speed and power. Employing quantum theory, doctors may make correlations and offer a diagnosis or therapy. Including its intelligence, quantum computing has the potential to transform current medicine completely. Quantum computing is not far behind when it comes to the detection and monitoring of illness. Chemotherapy is typically administered to cancer patients who do not learn the treatment results for several months. But that is altering now due to developments in quantum computing. Doctors can analyze enormous amounts of data concurrently in parallel, together with all possible combinations of that data, to identify the most accurate trends that characterize it.

It takes a while to fully comprehend how one medicine interacts with others in a combination. Given that quantum computing has the computer ability to simulate every scenario, it can dramatically reduce time. Precise clinical imaging and therapeutic procedures may be provided with the use of quantum computing. The viewing of single molecules is made possible by the incredibly accurate imaging produced by quantum imaging devices. A doctor can be helped by machine learning techniques and quantum computing when analyzing therapy outcomes. Quantum computing can aid in interpreting the treatment's results, and machine learning can assist in detecting anomalies in the human body. Light and dark areas can be seen on standard MRIs, and the radiologist must assess the problems. However, quantum imaging techniques can distinguish between various tissue types, enabling more accurate and thorough imaging.

## 7. Quantum Algorithms for Healthcare

Algorithmic medical research is projected to change due to the use of quantum computing in healthcare and life sciences. The US federal government has made it clear that it is dedicated to a future powered by quantum technology and is actively advancing this sector. The diagnosis of illnesses, the creation of medications, the development of methods for individualized medical therapies, and the analysis of medical imagery are some exciting uses of quantum technology. Additionally, quantum technology may significantly advance our comprehension of protein folding. Computational biologists have created excellent algorithms in recent years to simulate the form of proteins. These models help scientists better comprehend the body's natural systems by demonstrating how protein folding determines biocompatibility. Such algorithms still lack the accuracy required to make the expected advancements in customized treatment, and future quantum computers could alter that.

The amount of data in the healthcare sector is exploding, and healthcare costs are also growing. Big data, advanced machine learning, supercomputers, and cloud services are all used in cognitive computing to assist clinicians in identifying diseases early, enhancing therapy outcomes, and eventually lowering healthcare costs. Quantum algorithms offer the ability to tackle computationally complex issues with traditional computers by performing computations at extraordinarily high speeds. This holds great promise for extracting valuable, decision-level data from medical pictures.

To successfully exploit the potential of superposition, computer scientists create algorithms that can benefit from this condition. You are not alone if you find it challenging to understand all of this. Even the brightest scientists have trouble understanding these ideas, mainly because quantum theory is still primarily an abstract concept. Nowadays, a traditional computer can produce the strategy in a clinically realistic amount of time utilizing a small number of data points. Nevertheless, quantum-inspired algorithms will enable medical technologies to run all conceivable permutations concurrently, utilizing many data points and creating an ideal strategy more quickly.

Among the various varieties of quantum algorithms that might be crucial to the healthcare industry, the breadth of applications of quantum-enhanced machine learning techniques stands out. This is because we are moving closer to a time when the characteristics of health datasets, such as their frequent variability and inequitable distribution, may provide challenging computational challenges for existing AI. For instance, scientists have been exploring how to use quantum approaches, primarily processes with huge matrices, to accelerate the computationally expensive processes at the core of machine learning and AI modelling.

## 8. Quantum Simulation and Healthcare

Although quantum simulation is undoubtedly one of the most promising quantum technologies, it is a broad field constantly evolving and where significant breakthroughs occur frequently. Feynman recognized the fantastic nature of entanglement in 1982. He claimed that when the number of quantum objects rises, the size of their Hilbert space grows exponentially with the increase of entangled things, which was one of the shocking deviations from conventional thinking. An intriguing example is a situation when atoms replace electrons in a crystal, or a periodic arrangement, in a standing wave potential formed by laser beams. One, two, or three-dimensional simulations of events are feasible. Quantum statistics play a significant part in laser cooling approaches that allow one to have atoms with thermal energy that is low relative to the potential's magnitude. Therefore, the atoms' state—whether fermions or bosons—is crucial. In reality, by choosing the proper isotope of atoms, one may decide whether to have fermions or bosons. Atoms like lithium or potassium are frequently employed for that purpose. Lithium, for example, has two isotopes despite having three electrons. Due to the uniform distribution of nucleons and electrons, lithium 7 is a boson. Considering three plus six is an odd number, lithium six is a fermion. Bosonic atoms can be cooled down to a point where a Bose-Einstein condensate is formed while only experiencing little thermal excitation. It is possible to reduce the many-body system of exchanging atoms to its most basic form. With fermions, however, the thermal element is far more critical, and even the most acceptable cooling methods only result in a non-negligible percentage of the Fermi temperature.

Considering traditional approaches can't predict how proteins and other complicated systems would react to new medications, the creation of new pharmaceutical treatments today entails a lot of experimentation. We now have new chances to create customized quantum simulators that can be made to address these processes thanks to quantum technology. Professor Lodahl will serve as the director of the "Solid-State Quantum Simulators for Biochemistry" facility, also known as "Solid-Q," and will receive 60 million kroner for his research. The centre's efforts will be focused on implementing and combining two categories of quantum simulation hardware that can calculate the quantum mechanical properties of complex biomolecules.

The director of the other centre, "Quantum for Life," is Professor Matthias Christandl from the Department of Mathematical Sciences at UCPH. In order to explore intricate biological processes, this initiative intends to create mathematical techniques that may be used to the quantum simulation of proteins. By their very nature, healthcare systems are diverse, with each device using a distinct architecture, technology, and system software. From the perspective of delays and security risks, this variability impacts communication effectiveness.

By utilizing the concepts of quantum physics in novel ways to produce and process information, quantum simulations and computers are opening up transformational possibilities. It is anticipated that such calculations would positively impact a number of fields, from daily tasks to the discovery of new scientific theories. Such preliminary findings demonstrate that quantum computing and simulations could significantly speed up the deployment of new technologies that are urgently required to meet the growing demands for energy while protecting the environment. Numerous early-stage implementations of quantum computing and simulations have already been evidenced.

Treatment with drugs is required to cure the majority of diseases, preserve health, or stop future decline. Above all, medication is a highly important factor in the ageing population since a significant portion of hospital admissions in this age category are caused by adverse reactions brought on by incorrect drug administration and prescription. Medical professionals must thus use great caution in this area. Authors [34] proposes a graph-based system that will enable medical professionals to predict the potential negative effects that a specific drug could have on the wellbeing of an ageing person, based on the patient's history of drug use, the effects of the drug as explained, and the patient's physiological and genetic predictors.

In this article [35], the potential of quantum computing for health systems is examined. We investigate application areas where the increased processing speed offered by quantum computing might revolutionise current healthcare systems. We list the essential conditions for applying quantum computing to the healthcare sector. To find security flaws in conventional cryptography systems, the authors conducted a thorough analysis of quantum cryptography from the viewpoint of healthcare systems. We conclude by looking at present issues, their causes, and potential future research areas related to using quantum computing systems in healthcare. The latency [36] and energy conservation constraints for real-time health data collecting and processing can be satisfied by a quantum computing system. Based on the amazing phenomena of quantum physics and quantum mechanics, quantum computing is a novel form of computing. It is a fantastic fusion of computer science, information theory, arithmetic, and physics.

## 9. Comparative Analysis of Existing Quantum Computing-based Approaches for Healthcare

Table 2 shows the recent and important quantum computing integrated approaches for healthcare discussed in the literature.

**Table 2.** Comparative analysis of existing quantum computing-based approaches for healthcare

| Authors | Year | A | B | C | D | E | F | G | H | I | Proposed Approach | Future Directions |
|---|---|---|---|---|---|---|---|---|---|---|---|---|
| Ikeda [14] | 2018 | ✓ | ✓ | × | × | ✓ | ✓ | ✓ | × | × | Introduced the integration of blockchain and quantum computing. | Formal verification and experimentation work can be conducted to evaluate the feasibility. |
| Kiktenko et al. [12] | 2018 | ✓ | × | × | × | ✓ | ✓ | × | × | × | Introduced theoretical quantum, cryptography and blockchain aspects briefly. | Formal verification and experimentation work can be conducted to evaluate the feasibility. |
| Chuntang et al. [10] | 2019 | ✓ | × | × | ✓ | ✓ | × | × | × | × | Quantum, cryptography and blockchain aspects and | Practical implementation and formal verification can be extended. |
| | | | | | | | | | | | their integration are discussed. | |
| Justinia [38] | 2019 | ✓ | × | × | × | ✓ | × | ✓ | × | ✓ | Discussed the details of blockchain technology and its importance to healthcare. | Experimental work integrating blockchain and healthcare can be conducted. |
| Farouk et al. [37] | 2020 | ✓ | × | × | ✓ | ✓ | ✓ | ✓ | × | × | Theoretical discussions over the usage of blockchain technology in the large-scale healthcare sector. | This work can be extended to include a quantum computing-based network for healthcare systems. |
| Nandni and Jahnavi [17] | 2021 | ✓ | × | × | ✓ | ✓ | ✓ | ✓ | × | × | Discussed the importance of cryptography and blockchain integration. Here, the focus is drawn on the strongness of the digital signature. | This work can be extended to design quantum gates and circuits for cryptography primitives and protocols use for blockchain in real-time applications. |
| Kumar *et al. [13]* | 2021 | ✓ | ✓ | ✓ | ✓ | ✓ | ✓ | × | × | × | A detailed review of quantum drones and networks is conducted. Here, the integration of advanced technologies with quantum drones and networks is elaborated. | This work can be extended to include a mathematical model and the feasibility to design circuits for quantum networks. |
| Singh et al. [6] | 2021 | ✓ | ✓ | × | ✓ | ✓ | ✓ | ✓ | × | × | The importance of quantum computing for climate change is explored. | The feasibility to practically integrating quantum computing for environmental solutions can be worked upon. |
| Qu et al. [39] | 2022 | ✓ | × | | ✓ | ✓ | ✓ | ✓ | ✓ | ✓ | Quantum computing aspects for securing medical data are discussed. Here, medical data security while processing is focused. Further, the importance of data security for healthcare systems using quantum approaches is discussed. | The feasibility of practically securing healthcare applications and data using quantum computing-based approaches can be explored for all three (processing, transmission, storage) stages of data can be explored. |
| Chen et al. [40] | 2022 | × | ✓ | × | × | ✓ | × | ✓ | ✓ | ✓ | Security attributes controllable through | More cryptography primitives and protocols and their |
| | | | | | | | | | | | quantum aspects are discussed. | attributes can be discussed. |
| Azzaoui et al. [41] | 2022 | ✗ | ✓ | ✗ | ✗ | ✓ | ✓ | ✓ | ✓ | ✓ | Quantum support to cloud architecture, and its ecosystem are discussed to secure healthcare data in healthcare systems. | Designing quantum computing-based software and hardware architecture to secure cloud infrastructure for healthcare can be worked upon. |
| Charles [42] | 2022 | ✓ | ✗ | ✗ | ✗ | ✗ | ✗ | ✓ | ✓ | ✗ | The importance of blockchain in the present and futuristic applications (including healthcare) are discussed. | This work can be extended to include quantum computing aspects in blockchain-based systems for securing futuristic applications. |
| Grosu et al. [43] | 2022 | ✓ | ✗ | ✗ | ✗ | ✗ | ✗ | ✓ | ✓ | ✗ | This work shows the present and futuristic need for mobile applications and blockchain for the healthcare system. Here, an analysis of mobile devices is conducted in detail. | This work can be extended to discuss the importance of quantum computing for mobile devices or resource-constraint devices useful for healthcare applications and data. |
| Gupta et al. [44] | 2022 | ✗ | ✓ | ✗ | ✗ | ✓ | ✓ | ✓ | ✓ | ✗ | Here, quantum computing and securing the COVID-19 patient data and records are discussed. | A detailed architecture over quantum computing infrastructure specification can be explored for healthcare. |
| Sultana et al. [45] | 2022 | ✗ | ✓ | ✗ | ✗ | ✓ | ✓ | ✓ | ✓ | ✗ | Post-quantum cryptography primitives and protocols for security healthcare records are discussed. | The use of post-quantum cryptography for healthcare to avoid attacks on traditional security mechanisms can be explored in detail. |
| Pandey et al. [46] | 2022 | ✓ | ✗ | ✗ | ✗ | ✗ | ✗ | ✓ | ✓ | ✗ | The importance of securing blockchain using non-fungible tokens is explored. | The importance of quantum computing-integrated tokens for healthcare and other applications can be elaborated. |
| Ahmad et al. [47] | 2022 | ✓ | ✓ | ✗ | ✗ | ✓ | ✗ | ✗ | ✗ | ✓ | Blockchain for a healthcare system with | The feasibility to improve the |
| | | | | | | | | | | | a detailed framework is elaborated. | scalability of a blockchain network for healthcare can be explored. |
| Zhu et al. [48] | 2022 | ✗ | ✓ | ✗ | ✓ | ✓ | ✗ | ✓ | ✗ | ✓ | A Quantum computing-supported encryption scheme for video conferences is discussed. | This work can be extended to include scenarios where patients and doctors can use this facility to securely exchange medical data and associated treatment. |
| Trenfield et al. [49] | 2022 | ✗ | ✓ | ✗ | ✓ | ✓ | ✗ | ✓ | ✗ | ✓ | The use of digital technologies in different scenarios for pharmacy is explored and discussed. | The evaluation (mathematically or practically) can be conducted for proposed scenarios. |
| Laxminarayana et al. [50] | 2022 | ✗ | ✓ | ✗ | ✓ | ✓ | ✗ | ✓ | ✗ | ✓ | Here, the importance of quantum computing in analyzing medical data using machine and deep learning approaches are explored and analyzed. | This work can be explored for more medical records/datasets to analyze and compare the performance with and without quantum computing. |
| Li et al. [51] | 2022 | ✗ | ✓ | ✗ | ✗ | ✓ | ✗ | ✓ | ✗ | ✓ | In this work, a blockchain-based healthcare system is designed to secure data storage and accessibility. In security, a searchable-encryption mechanism is used for the same. | The evaluation of blockchain networks over their scalability, performance and network security can be worked upon for their acceptability on a large scale. |
| Tchagna et al. [52] | 2022 | ✓ | ✓ | ✗ | ✗ | ✓ | ✗ | ✓ | ✓ | ✓ | A case study to secure medical data for healthcare in a smart city is prepared and discussed. | The practical feasibility to integrate fast computing (like quantum computing) can be explored. |
| Mahajan et al. [24] | 2022 | ✗ | ✓ | ✗ | ✗ | ✓ | ✗ | ✓ | ✗ | ✓ | Healthcare, quantum computing and cloud infrastructure and their integrated advantages are explored and discussed in detail. | This work can be extended to include security primitives and protocols that integrate three technologies for ensuring full-proof security. |
| Gupta et al. [27] | 2022 | × | ✓ | × | × | ✓ | × | ✓ | × | ✓ | In this work, quantum computing supported blockchain for efficient authentication in the internet of vehicles is proposed. | The authentication mechanism evaluation against various attacks can be formally conducted and evaluated. |
| Malviya and Sundram [36] | 2022 | × | ✓ | × | × | ✓ | × | ✓ | × | ✓ | The importance of quantum computing in the smart healthcare system is discussed with its pros and cons. | The analysis can be extended to have in-depth tools and techniques-based analysis for identifying and generating efficient healthcare scenarios. |

A: quantum blockchain, B: quantum integrated technologies (quantum drones, quantum satellites, quantum AI/ML), C: photonic quantum computing, D: quantum gates and circuits, E: quantum algorithms, F: quantum simulation, G: healthcare records, H: IoT-based services, I: Other advanced technologies (cloud computing, serverless computing, cybersecurity, forensics, metaverse and others).

**Pros of Quantum Computing-based Approaches for Healthcare:** The use of quantum computing in healthcare can (i) improve the security of the healthcare system again various real-time attacks, (ii) faster the data processing and availability for authenticated users, (iii) improve the security and enhances the healthcare system for every stakeholder, (iv) Smart healthcare scenarios for improving patient handling and patient-centric system design can be easily achievable, and (v) Quantum blockchain-based scenarios can support in few challenge generation, security of challenges for miners, improved security for authenticated users, protection again attacks and many more.

**Cons of Quantum Computing-based Approaches for Healthcare:** The major challenges in the use of quantum computing for healthcare systems include: (i) large-scale use of quantum computing may not be environmentally friendly. (ii) feasibility to arrange quantum computing resources for the healthcare system in real-scenario is a challenging task, especially in developing countries. (iii) Infrastructure to support quantum computing for real-time applications (especially for healthcare) is a daunting task. It is a costly solution as well. Although tools (like Qiskit, Silq etc.) are available for computing for small-scale solutions. To achieve large-scale integration, more Qibit-support machines are required.

## 10. Conclusion and Future Directions

The use of blockchain technology in the fight against the COVID-19 epidemic as well as in other healthcare systems of a similar kind, has shown to be highly advantageous. The most up-to-date medical care systems include sensor data, which may be used to monitor patients while also protecting their privacy and the confidentiality of their medical records. This technology can store enormous quantities of data in an organized and safe way, enabling users to access that data whenever they wish. This puts it in a class above all other technologies, which puts them in a lower class. Because they may be obtained promptly and conveniently whenever and wherever required, they can be used whenever and wherever it is essential. Patients may be traced down and found using quantum blockchain technology and network in a couple of minutes. In a quantum blockchain, it is feasible to conceal data while ensuring that it is secure and not too complicated to access. It's possible that using technologies like quantum computing and blockchain may make it possible to handle patient data more quickly while maintaining its integrity. Both blockchain technology and quantum technologies have been the subject of a great deal of conjecture. This investigation aimed to determine whether or not there are presently operational medical uses for either of these technologies. Research is being conducted in this field on various technologies, including quantum physics, blockchain, ratification intelligence, machine learning, and drones.

**Conflict of Interest Statement**

On behalf of all authors, the corresponding author states that there is no conflict of interest.

**Data Availability statement**

My manuscript has no associated data.

**Funding**
This research received no specific grant from any funding agency in the public, commercial, or not-for-profit sectors.'

**Acknowledgement**
I would like to express my very great appreciation to Dr K. Rajalakshmi and all unknown reviewers for her valuable and constructive suggestions during the exploration of this research work. Her willingness to give her time so generously has been very much appreciated.